Relativistic effects for the superheavy reaction Og +2Ts2 → Og(Ts)$_4$ (T$_d$ or D$_{4h}$): Dramatic relativistic effects for atomization energy of superheavy Oganesson tetratennesside Og (Ts)$_4$ and prediction of the existence of tetrahedral Og (Ts)$_4$

Gulzari L Malli [a]

Department of Chemistry, Simon Fraser University,

8888 University Dr., Burnaby, BC; Canada V5A 1S6

Martin Siegert, [b]

Research Computing, IT Services, Simon Fraser University

8888 University Dr., Burnaby, BC; Canada V5A 1S6

Luiz Guilherme M. de Macedo [c]

Universidade Federal de São João del-Rei,UFSJ

Campus Centro-Oeste Dona Lindu - CCO, UFSJ

Bairro Chanadour - Divinópolis, MG - Brazil 35.501-296

Walter Loveland [d]

Department of Chemistry, Oregon State University

Corvallis, Oregon, OR 97331-4003, USA

[a] Corresponding author email: malli@sfu.ca

[b] email:siegert@sfu.ca

[c] email:lgm@ufsj.edu.br

[d] email:lovelanw@onid.orst.edu

**Abstract** Our all−electron fully relativistic Dirac−Fock (DF) and nonrelativistic (NR) Hartree-Fock (HF) SCF molecular calculations for the superheavy tetrahedral (Td) oganesson tetratennesside OgTs$_4$ predict atomization energy (A$_e$) of 7.45 and -11.21 eV, respectively. Our DF and NR calculations, however for the square planar (D$_{4h}$) OgTs$_4$ predict atomization energy (A$_e$) of 6.34 and -8.56, eV respectively. There are *dramatic relativistic effects* for the atomization energy of T$_d$ and D$_{4h}$ OgTs$_4$ of ~ 18.65 eV and ~ 14.90 eV, respectively. Whereas our DF calculations predict the T$_d$ OgTs$_4$ *to be more stable than the* D$_{4h}$ OgTs$_4$ *by ~ 1.10 eV*, our *NR calculations predict the* D$_{4h}$ OgTs$_4$ to be *more stable than the* T$_d$ OgTs$_4$ by ~2.65 eV. Our NR calculations predict both the T$_d$ and D$_{4h}$ OgTs4 to be *unbound* by 11.21 and 8.56, eV, respectively. However, our relativistic DF calculations predict both the T$_d$ and D$_{4h}$ OgTs4 *to be bound* by 7.45 and 6.34,eV respectively and so the relativistic treatment is mandatory for bonding and binding in the pentatomic superheavy system with 586 electrons involving the two heaviest SHE Ts and Og.

1 Introduction

During the last decade, there have been numerous investigations of the superheavy elements (SHE) with Z > 103 [1-9]. Recently [9] four superheavy elements (SHE) have been placed in the 7th row of the periodic table including the two heaviest SHE Tennessine Ts (Z=117) and Oganesson Og (Z=118). This is a landmark event for the scientists working in this area of research and should lead to a renewed interest in the experimental as well as theoretical investigation of the physical and chemical aspects of these heaviest SHE. It is well-recognized that there are serious problems with the experimental studies of the SHE, due to small production cross-section, extra short lifetime, and the access of one atom at a time for chemical study, etc. However, except for a few diatomics of Ts and Og, there are hardly any ab initio all-electron relativistic and non-relativistic calculations especially for systems of the two heaviest SHE Og and Ts. It is well-known that relativistic effects may be quite pronounced for atomic and molecular systems of SHE, and in the investigation of their electronic structure, bonding, chemical behavior, etc, Schrodinger's nonrelativistic treatment may be inadequate while Dirac's relativistic treatment for many-electron systems may be more appropriate for such systems. Dirac-Fock (DF) SCF theory for molecules was developed by Malli and Oreg [10] in 1975 and has been used extensively [11-18] to investigate the effects of relativity in the chemistry of heavy actinides, and superheavy elements. Recently we have investigated [11-18] the effects of relativity on the electronic structure and bonding of numerous systems of heavy and superheavy elements (SHE). Our goal in this paper is to investigate the relativistic effects on the chemistry of Oganesson and especially for the atomization energy of Og (Ts)$_4$ and energy of reaction for the superheavy reaction:

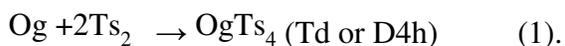

Og +2Ts$_2$ → OgTs$_4$ (Td or D4h)        (1).

**2 Methodology**

Dirac-Fock (DF) SCF methodology for molecules [10-18] has been used quite extensively to incorporate effects of relativity on the electronic structure and chemical bonding for molecular systems of heavy and superheavy elements(SHE) and the reader is referred to these papers for further computational details. We use the dyall.cv4z basis for both the atoms Og and Ts for all our calculations which are performed with DIRAC [19] code (hereafter referred to as code). However, for the NR HF calculations on Td and D4h OgTs$_4$, we use the dyall.cv2z basis. We use the gaussian nuclear model for Og and Ts and the small component wavefunctions of Og and Ts are obtained from their large component using the kinetic balance [20] as implemented in the code.

## 3 DiracFock and NR Hartree-Fock calculations for tetrahedral (Td) and square planar ( D4h) Og(Ts)$_4$ and Ts$_2$

Neither *ab initio* all-electron 4-component relativistic DF nor the corresponding NR Hartree-Fock calculations are available for the superheavy OgTs$_4$ with five superheavy atoms and 586 electrons. Needless to say that DF and NR HF calculations for OgTs$_4$ would be gargantuan and would require a state-of-the-art supercomputer facility. Most of our calculations were performed using the DIRAC code [19]. All-electron Dirac-Fock (DF), and the corresponding nonrelativistic (NR) Hartree-Fock calculations for the OgTs$_4$ were performed with the Dirac [19] code using the dyall.cv4z basis for Og and Ts, which are available within the Dirac website [19]. The total number of primitive gaussians used in our calculations is 8220 with 2550 L (large component) and 5670 S (small components) and the large component basis set for Og and Ts is 35s 35p 24d 16f 3g 1h. The kinetic balance [20] constraints as well as, the gaussian nuclear model as implemented in the code are employed in all of our calculations. Our calculated total DF energies using the above-mentioned basis and atomic masses used in the DIRAC code for Og and Ts are –54807.9332 and -53494.0590 au, respectively. The LEVY-LEBLOND option was used for all the NR calculations and our calculated total NR HF energies for Og and Ts are -46320.4278, and -45396.6978 au, respectively. All of our DF calculations for the atoms Og and Ts, the diatomic Ts$_2$, and the polyatomic OgTs$_4$ were carried out using the above-mentioned appropriate basis with the Dirac code [19].

However, the NR HF calculations for the Td and D4h OgTs$_4$ were performed using the dyall.cv2z basis. Geometry optimization for the Td and D4h OgTs$_4$ and the diatomic Ts$_2$ were carried out automatically using the Dirac code [19]. Mulliken population analysis [21] as implemented in the Dirac code [19] was carried out to obtain the charges on the Og and Ts atoms; however, it should be stressed, that this analysis is highly basis set dependent [22] and its results should be used with due caution. The results of our NR and DF SCF calculations for OgTs$_4$ are collected in Table 1.

## 4. Relativistic effects for the atomization energy of OgTs$_4$ and energy of the reaction Og + 2Ts$_2$ → Og (Ts)$_4$ (Td or D4h)

The atomization energy (dissociation into one atom of Og (g) plus 4 atoms of Ts (g)) $A_e$ for OgTs$_4$ (T$_d$.) calculated with our relativistic DF and NR HF SCF wavefunctions is 7.45 and -11.21 eV, respectively. Mulliken [21] population analysis of our relativistic DF and NR HF wavefunctions for the Td OgTs$_4$ yields charge of + 0.64 and +1.64 on Og, respectively; and therefore our HF NR wavefunction predicts Td OgTs4 to be *more ionic* as compared to that predicted with our relativistic wavefunction. It should be mentioned, however, that Mulliken's population analysis results are highly basis set dependant [22] and should be treated with due caution. The results of our calculations of atomization energy at the optimized bond lengths are presented in Table I.

# 5. Relativistic effects for electronic structure, bonding, and molecular spinor energy levels of the $e_1$ symmetry of tetrahedral (Td) OgTs$_4$

There are 293 $e_1$ doubly occupied molecular spinors (MS) in the relativistic ground state closed-shell electron configuration of the tetrahedral (Td) OgTs$_4$ with 586 electrons. We shall discuss here the relativistic 293 MS's labeled as $1e_1$ to 293 $e_1$ in the ascending order of spinor energy. The lowest energy MS $1e_1$, with the energy of -8185.8357 au consists of 0.76 Og(1s) plus 0.24 of small component (SC) of Og(1s), which is the same as the atomic contribution to the higher energy 2s…6d ASs of Og and Ts decreases progressively from ~0.24 to 0.001. The next 4-fold degenerate Ms's 2 $e_1$ −5 $e_1$ with the energy of -7988.4484 au consist of 0.77 Ts (1s) plus 0.23 SC, which equals the atomic spinor (AS) Ts(1s). The MS 6 $e_1$ with the energy of -1718.9800 au consists of 0.94 Og(2s) plus 0.06 SC and equals the AS Og(2s). The next MS $7e_1$ of energy -1681.8118 au consists of 0.94 Og(2p) plus 0.06 SC and equals the AS Og(2p$_-$).

The degenerate MS's (8 $e_1$ - 11 $e_1$) with the energy of -1669.6851 au, consist of 0.94 Ts(2s) plus 0.06 SC, which equals the AS Ts(2s). It can be seen that the MS's ($1e_1$ - 11 $e_1$) consist of the core-like 1s-2p AS of Og and Ts and all these MS's are nonbonding. To continue, the MS's ( 12 $e_1$ - 285 $e_1$ ) consist of the core-like atomic spinors up to the higher energy valence AS Ts(7p$_+$) and Og (7p$_+$). This implies that all the first 285 MS's ($1e_1$ - 285 $e_1$ ) of Td OgTs$_4$ are core-like atomic spinors of the two atoms Ts and Og ( up to the valence Ts(7p$_-$) and Og(7p$_-$) and all the 285 MS's are non-bonding. The doubly degenerate MS's (286$e_1$ - 287$e_1$) with the energy of -0.3724 au consist of the valence atomic spinors 0.73 Og (7p$_+$) and 0.24 Ts (7p$_+$) and are bonding MS's. The next degenerate MS's (288$e_1$ - 291$e_1$) with the energy of -0.2516 au, consist of 0.97 Ts(7p$_+$), and are nonbonding. The MS 292$e_1$ with the energy of -0.2316 au consists of 0.90 Ts(7$p_+$), 0.07 Og(7s) and 0.02 Ts(7s) and is an antibonding MS. Finally, the highest occupied molecular spinor(HOMS) is the MS 293 $e_1$ with the energy of -0.2039 au consists of 0.95 Ts(7$p_+$) and 0.03 Og (7$p_+$) AS's and is antibonding.

Our calculated energies of the titled reactions at the optimized geometry for Td OgTs$_4$ at the DF and NR levels of theory are 0.95 and 13.23, eV, respectively. However, the DF and NR energies of the titled reaction leading to D4h OgTs$_4$ are 2.06 and 10.58 eV, respectively.

## 6. Non-relativistic electronic structure, Bonding, and molecular orbital energy levels of tetrahedral (Td) OgTs$_4$

There are 293 $e_1$ doubly occupied molecular orbitals (MO) in the nonrelativistic (NR) Hartree-Fock(HF) ground state closed-shell electron configuration of the tetrahedral (Td) OgTs$_4$ with 586 electrons. We shall discuss here the NR 293 MO's labeled as 1$e_1$ to 293 $e_1$ in the ascending order of energy.

The lowest energy MO 1$e_1$, with the energy of -62221.1434 au consists of 0.999 Og(1s). The next (2$e_1$ - 5 $e_1$ ) MO's are 4-fold degenerate with the energy of -6112.3810 au and consist of 1.000 Ts(1s) AO. The MO 6$e_1$ with an energy of -1144.6459 au consists of 0.999 Og(2s) and all the (1$e_1$ - 6 $e_1$ ) MO's are nonbonding as these consist of inner-core Og(1s), Ts(1s), Og(2s) AO's of the Og and Ts atoms. It is astonishing ( similar to the case of the relativistic energy levels of OgTs$_4$ discussed above in section 5 that all the NR MO's ( 1$e_1$ -275$e_1$) consist of pure AO's Ts(1s)…Ts(6d) and Og(1s) … Og(6d) and all the 275 MO's are nonbonding. However, the next MO $276e_1$ with the energy of -0.9098 au consists of 0.90 Og(7s), 0.06Ts (7s, and 0.02 Ts(7p) and is very strongly bonding. The next triply degenerate MO's ($277 e_1$-$279e_1$) with the energy of -0.7110 au consist of 0.88 Ts(7s) and 0.11 Og(7p) and are bonding. The MO $280e_1$ with the energy of -0.6789 au consists of 0.94 Ts(7s) and 0.06 Og(7s) and is weakly bonding. The remaining MO's ($281e_1$- $293e_1$) with energy from -0.5198 au to -0.2875 au consist of the valence AO's Ts(7s), Ts(7p), Og(7s), Og(7p), and all the MO's are weakly bonding except the HOMO 293 e1 which is antibonding. This completes the bonding and NR MO energy level analysis for the NR Td $OgTs_4$

**7. Conclusion** We have performed the first all-electron fully relativistic Dirac-Fock and the corresponding nonrelativistic calculations for $OgTs_4$ a pentatomic of five superheavy elements with 586 electrons and a summary of our major conclusions is as follows:

(1) Our relativistic DF SCF calculations predict the superheavy tetrahedral (Td) Og $(Ts)_4$ to be *bound*, with the calculated DF atomization energy of 7.45 eV at the optimized Og-Ts bond distance of 3.43 Å; however, our DF SCF calculation predicts the superheavy square planar (D4h) Og $(Ts)_4$ to be *bound* with the calculated DF atomization energy of 6.34 eV at the optimized bond distance Og-Ts of 3.23 Å.

(2) Relativistic effects of ~18.65 eV to the atomization energy are *dramatic* and must be calculated using *ab initio* DF SCF or better methodology

(3) Our *ab initio* all-electron fully relativistic DF and nonrelativistic Hartree-Fock (NR) calculations *predict* the DF and NR HF energy of the titled reaction Og +$3Ts_2$ → $OgTs_4$ leading to Td $OgTs_4$ as 0.95 and 13.23, respectively. The relativistic effects of ~- 14 eV to the energy of the titled reaction are very significant and proper and rigorous relativistic treatment is mandatory for calculating the relativistic effects in chemical reactions involving superheavy systems

(4) However, Our *ab initio* all-electron fully relativistic  DF  and nonrelativistic Hartree-Fock (NR)  calculations *predict* the  DF  and NR  HF energy of the titled reaction (Og +3Ts$_2$ → OgTs$_4$  (D4h) (leading to D4h OgTs4)   as 2.06 and 10.58  eV, respectively. The relativistic effects of ~ -12.65 eV to the energy of the titled reaction (leading to D4h OgTs4) are very significant and proper and rigorous relativistic treatment is mandatory for the calculation of the relativistic effects in chemical reactions involving superheavy systems.

(5) There are very large relativistic corrections to the binding energies of the MOs, especially, the inner core orbitals of Og (Ts)$_4$. Moreover, very large S–O splitting is calculated for the core MOs that consist of the inner (core) p, d, and f  AOs of  the Og as well as the Ts atoms as expected.

(6) The 1s...7s AS's of the Og atom as well as the 1s… 7s  AS's of the four Ts ligands, and their associated electrons are not involved in bonding in Og(Ts)$_4$, since they remain as if in pure AS's of Og and Ts atoms. Therefore, these core electrons could be treated in molecular calculations on compounds of the superheavy elements (SHE) Og and Ts, using appropriate frozen core or pseudopotential approximations with tremendous savings in computational cost.

(7) The predicted DF and NR Og-Ts bond distances for  the T$_d$  OgTs$_4$ are 3.43 and 3.45 Å, respectively and therefore the relativistic effects are not significant for calculation of the bond distances of superheavy systems like  T$_d$ OgTs$_4$. However  the predicted  DF and NR Og-Ts bond distances for the square planar (D4h) OgTs4 are 3.23 and 3.46 Å ,respectively and therefore a correct treatment of relativistic effects is necessary to assess the relativistic effects on bond distances in such systemseffec

In conclusion, *ab initio fully relativistic all-electron* Dirac-Fock SCF calculations for molecular systems of SHE with about 600 electrons are no longer the *bottlenecks* of relativistic quantum chemistry of SHE.

**Acknowledgments**: This research used in part resources of the National Energy Research Scientific Computing Center (NERSC), which is supported by the Office of Science of the U.S.Department of Energy under Contract No.DE-AC03-76SF00098. Part of our extensive calculations was carried out using the Westgrid computing resources which are gratefully acknowledged.

References:

1. Yu.Ognaessian and S.N.Dmitrieve, Russ.Chem.Rev, **85**, 901(2016)

2. M. Munzenberg, The European Physical Journal Conferences, **182:**02091 (2018), https://doi.org/10.1051/eppjconf/20181820291

3. *The Chemistry of Superheavy Elements*, Matthias Schaedel and Dawn Shaughnessy (eds), Springer Verlag, Berlin Heidelberg, 2014

4. Yu.Ognaessian and S.N.Dmitrieve, Russ.Chem.Rev, **85**, 901(2016)

5. Josh Fischman, Scientific American, January 4, 2016

6. H. W. Gaggeler, Radiochim. Acta 99, 503 (2011)

7. Yu.Ts.Oganessian, Phys.Rev.Letters.**104**, 142502 (2010)

8. A.Turler and V.Pershina, Chem. Rev. **113, 1237**(2013)

9. P.J.Karol, R.C.Barber, B.M.Sherrill, E.Vardaci and T.Yamazaki, Pure App.Chem **88**,155 (2015)

10. G.L Malli and J. Oreg, J.Chem.Phys. **63**, 830 (1975).

11. G.L.Malli, in *Proceedings of the Robert A. Welch Foundation, 41st Conference on Chemical Research THE TRANSACTINIDE ELEMENTS*, pp 197-228, Houston Texas, October 27-28, 1997.

12 G.L. Malli, .and J. Styszynski*, J.Chem.Phys. **104**, 1012, (1996)

13. G.L.Malli, J.Chem.Phys. **109**, 4448 (1998)

14. G.L.Malli in *Fundamental World of Quantum Chemistry* Vol III, edited by E.J.Brandas and E.S.Kryachko. (Kluwer Academic Press, Dordrecht, 2004), pp. 323-363.

15. G.L.Malli, Theor Chem. Acc. **118**, 473 (2007).

16. G.L. Malli, .and J. Styszynski*, J.Chem.Phys. **101**, 10736 (1994)

17 G.L. Malli, J.Chem.Phys. **144**, 194301 (2016)

18. G.L. Malli, J.Chem.Phys. **101**, 6829 (1994)

19. DIRAC, a relativistic ab initio electronic structure program, Release DIRAC12 (2012), written by H. J. Aa. Jensen, R. Bast, T. Saue, and L. Visscher, with contributions from V. Bakken, K. G. Dyall, S. Dubillard, U. Ekstroem, E. Eliav, T. Enevoldsen, T. Fleig, O. Fossgaard, A. S. P. Gomes, T. Helgaker, J. K. Laerdahl, Y. S. Lee, J. Henriksson, M. Ilias, Ch. R. Jacob, S. Knecht, S. Komorovsky, O. Kullie, C. V. Larsen, H. S. Nataraj, P. Norman, G. Olejniczak, J. Olsen, Y. C. Park, J. K. Pedersen, M. Pernpointner, K. Ruud, P. Salek,B. Schimmelpfennig, J. Sikkema, A. J. Thorvaldsen, J. Thyssen, J. van Stralen, S. Villaume, O. Visser, T. Winther, and S. Yamamoto (see http://www.diracprogram.org).

20. R. E. Stanton and S. Havriliak, J.Chem.Phys. **82**, 1910 (1984)

21. R. S. Mulliken, J.Chem.Phys. **23**, 1833 (1955).

22. K. R. Roby, Mol. Phys. **47**, 81 (1974).

**Table I.** Calculated total nonrelativistic (NR) HF and relativistic Dirac-Fock (DF) energies (in au) for $OgTs_4$ ($T_d$) and $OgTs_4$ (D4h) at the optimized bond distance ($R^X$ in Å), total energy ($E^X$ in au), atomization energy ($Ae^X$ in eV) for $OgTs_4$ ($T_d$) and $OgTs_4$ (D4h) and energy for the reaction ($\Delta E^X$ in eV) for the reaction $Og + 2Ts_2 \rightarrow OgTs_4$ as predicted with our NR and DF calculations

|  | $OgTs_4$ (Td) | $OgTs_4$ (D4h) |
| --- | --- | --- |
| $E^{NR}$ | -227906.8005 | -227906.8980 |
| $E^{DF}$ | -268784.4429 | -268784.4022 |
| $Ae^{NR}$ | -11.21 | -8.56 |
| $Ae^{DF}$ | 7.45 | 6.34 |
| $\Delta E^{DF}$ | 0.95 | 2.06 |
| $\Delta E^{NR}$ | 13.23 | 10.58 |
| $R_{Og-Ts}^{DF}$ | 3.43 [a] | 3.23 |
| $R_{Og-Ts}^{NR}$ | 3.45 | 3.46` |

[a] distorted Td with all non-tetrahedral angles